\documentclass[twocolumn,aps,amsmath,amssymb,prl]{revtex4}

\usepackage{graphicx} 
\usepackage{dcolumn}  
\usepackage{bm}       
\usepackage{ogonek}

\begin{document}
\pagestyle{empty}

\parskip=0cm

\noindent
{\large\bf Comment on ``Ultrametricity in the Edwards-Anderson Model''}
\smallskip
\noindent

In a recent interesting Letter Contucci {\it et al.}\cite{Ultra} have
investigated several properties of the three-dimensional (3d) Edwards-Anderson
(EA) Ising spin glass. They claim to have found strong numerical evidence 
for the presence of a complex ultrametric structure similar to the one 
described by the celebrated replica symmetry breaking (RSB) solution of 
the mean field model\cite{MPV}. Considering three spin configurations at
thermal equilibrium and their mutual link overlaps $(Q_{12},Q_{23},Q_{31})$, 
ultrametricity states that only equilateral and isosceles triangles of sides
$(Q_{12},Q_{23},Q_{31})$ are observed. As a consequence, if $u={\text {min}}
(Q_{12},Q_{23},Q_{31})$, $v={\text {med}}(Q_{12},Q_{23},Q_{31})$ and 
$z={\text {max}}(Q_{12},Q_{23},Q_{31})$, the following identities 
hold for the distributions of $x=v-u$ and $y=z-v$:
\begin{eqnarray}
\tilde{\rho}(x) &=& \delta(x),\label{um1}\\
\tilde{\rho}(y) &=& \frac 14 \delta(y) + \frac 32 \theta(y) \int_y^1 P(a)P(a-y)da,
\label{um2}
\end{eqnarray}
where $P(Q)$ is the probability distribution of the link overlap.

The authors of \cite{Ultra} state that the droplet model, where only one state
and its reversal symmetric exist in the thermodynamic limit, cannot satisfy
non-trivial ultrametricity because its $P(Q)$ is trivial so that only
equilateral triangles will be observed for the link overlap.  This is not quite
true, as its $P(Q)$ (for both link and site overlap) for finite sizes and
temperatures can be far from showing a trivial structure.  In taking {\it first}
the thermodynamic limit and {\it then} testing the relations in
Eqs.\eqref{um1} and \eqref{um2},
an implicit choice of limits is made which has already been shown to lead to a
wrong interpretation in Ref.~\cite{BMMK} for the so-called Guerra
parameter. In the following we illustrate that some caution is needed before 
dismissing other pictures by the approximate numerical verification of
Eqs.\eqref{um1} and \eqref{um2}.

A first example of this was given by the analytical study of Bray and Moore
\cite{BM1d} who showed that in a one-dimensional spin glass, for sizes below
the equilibrium length scale, the distribution of three overlaps follows a
non-trivial relation which is very similar to the ultrametric relation
Eq.(10) of Ref.~\cite{Ultra}.

\begin{figure}[t]
  \centering
  \includegraphics[width=0.9\columnwidth]{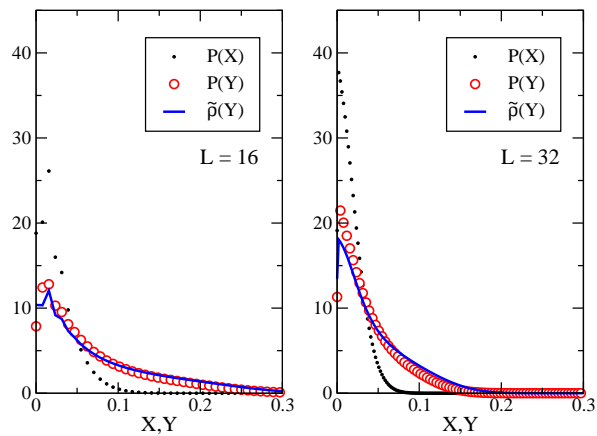}
  \caption{Empirical distributions ${\cal P}(X)$ and ${\cal P}(Y)$ for
    $X=Q_{med}-Q_{min}$ and $Y=Q_{max}-Q_{med}$ for the two system sizes
    ($L=16$ and $L=32$) at temperature $T=0.2$ for the two-dimensional (2d) Edwards-Anderson
    model. $\tilde\rho(Y)$ shows the distribution of $Y$ computed via
    Eq.(\ref{um2}) using experimental data for ${\cal P}(Q)$ and approximating
    the delta function with the histogram $X$. This plot shows the same
    quantities as Fig.~2 of Ref.~\cite{Ultra}, however, for the 2d model.}
\label{fig}
\end{figure}

A second example we obtain by performing Monte Carlo simulations of the
two-dimensional (2d) EA model with Gaussian couplings. We compute the $P(Q)$ 
for different sizes and test if Eqs.\eqref{um1} and \eqref{um2} are approximately verified. 
In doing so, we are in fact able to reproduce in the 2d model the same
behavior shown in Fig.~2 of Ref.~\cite{Ultra} for the 3d model with similar
precision (see our Fig.~\ref{fig}). In \cite{Ultra},
these data are used as a strong hint for the presence of a spin glass phase
with an ultrametric structure while we obtain almost identical results in the
2d model where it is widely accepted that (i) there is no spin glass phase at
any finite temperature, and that (ii) the transition is well-described by the
droplet picture.

To conclude, we believe that although the data published in \cite{Ultra} may
be compatible with the presence of an ultrametric structure, they are, however,
not sufficient to dismiss the possibility that other models as, e.g., the 
droplet model might apply to the 3d EA spin glass. We also want to
emphasize the need of comparing with simple models in order to validate 
the conclusions reached in large-scale numerical experiments, especially 
in the context of spin glasses where simulations and their interpretation 
are known to be difficult.

\bigskip
\noindent
Thomas J\"org$^{1,2}$ and Florent Krz\k{a}ka{\l}a$^3$\\ \vspace{-0.2cm}
{\small
  \begin{tabbing}
    \indent$^1$ \=LPTMS, UMR 8626 CNRS
    et Universit\a'{e} de Paris-Sud,\\
    \>91405 Orsay Cedex, France\\
    \indent$^2$ Equipe TAO - INRIA Futurs,\\
    \>91405 Orsay Cedex, France\\
    \indent$^3$ PCT, UMR Gulliver CNRS-ESPCI 7083,\\
    \>10 rue Vauquelin, 75231 Paris Cedex 05, France\\
  \end{tabbing}
}\vskip-0.1cm
\noindent
Date: 6 September, 2007

\end{document}